\documentclass[twocolumn,prl,showpacs,preprintnumbers,amsmath,amssymb, superscriptaddress]{revtex4-1}
\pdfoutput=1

\usepackage{graphicx}
\usepackage{dcolumn}
\usepackage{bm}
\usepackage{graphicx}
\usepackage{xspace}
\usepackage{multirow}
\usepackage{natbib}
\usepackage{color}
\usepackage{CJK}

\begin{document}

\title{Observation of momentum space semi-localization in Si-doped $\beta$-Ga$_2$O$_3$}

\author{P. Richard}\email{p.richard@iphy.ac.cn} 
\affiliation{Beijing National Laboratory for Condensed Matter Physics, and Institute of Physics, Chinese Academy of Sciences, Beijing 100190, China}
\author{T. Sato}
\affiliation{Department of Physics, Tohoku University, Sendai 980-8578, Japan}
\author{S. Souma}
\affiliation{WPI Research Center, Advanced Institute for Materials Research, Tohoku University, Sendai 980-8577, Japan}
\author{K. Nakayama}
\affiliation{Department of Physics, Tohoku University, Sendai 980-8578, Japan}
\author{H. W. Liu}
\affiliation{Beijing National Laboratory for Condensed Matter Physics, and Institute of Physics, Chinese Academy of Sciences, Beijing 100190, China}
\author{K. Iwaya}
\affiliation{WPI Research Center, Advanced Institute for Materials Research, Tohoku University, Sendai 980-8577, Japan}
\author{T. Hitosugi}
\affiliation{WPI Research Center, Advanced Institute for Materials Research, Tohoku University, Sendai 980-8577, Japan}
\author{H. Aida}
\affiliation{Namiki Precision Jewel Co., Ltd., Tokyo 123-8511, Japan}
\author{H. Ding} 
\affiliation{Beijing National Laboratory for Condensed Matter Physics, and Institute of Physics, Chinese Academy of Sciences, Beijing 100190, China}
\author{T. Takahashi}
\affiliation{Department of Physics, Tohoku University, Sendai 980-8578, Japan}
\affiliation{WPI Research Center, Advanced Institute for Materials Research, Tohoku University, Sendai 980-8577, Japan}

\begin{abstract}
We performed an angle-resolved photoemission spectroscopy study of Si-doped $\beta$-Ga$_2$O$_3$. We observed very small photoemission intensity near the Fermi level corresponding to non-dispersive states assigned to Si impurities. We show evidence for a quantization of these states that is accompanied by a confinement in the momentum space consistent with a real-space finite confinement observed in a previous scanning tunneling microscopy study. Our results suggest that this semi-localization in the conjugate spaces plays a crucial role in the electronic conduction of this material.
\end{abstract}



\maketitle

Its transparency and relatively high conductivity compared to other semiconductors, even at low temperature ($T$), promote the wide gap semiconductor $\beta$-Ga$_2$O$_3$ as an attractive candidate for optoelectronic applications \cite{UedaAPL97}. Despite general enthusiasm for this material, there is no commonly accepted picture to explain its conductivity, which calls for a complete characterization of the electronic states near the Fermi level ($E_F$) of this material. A recent scanning tunneling microscopy (STM) investigation of Si-doped $\beta$-Ga$_2$O$_3$ evidenced the presence of Si impurity donor states very near $E_F$ \cite{IwayaAPL2011}. From their spatial extension $\Delta r$, the authors conjectured that conductivity likely occurs from hopping between nearby impurities. However, STM is a real ($r$) space measurement that cannot image directly how electrons propagate in materials. As a natural  complement, angle-resolved photoemission spectroscopy (ARPES) is a powerful tool to access directly the electronic band structure of materials in the momentum ($k$) space that can be used to address this problem. Unfortunately, previous ARPES studies of $\beta$-Ga$_2$O$_3$ did not report any state near $E_F$ \cite{LovejoyAPL2009,JanowitzNJP13}.  

Here we report an ARPES characterization of the electronic states at low-energy ($E$) in the $k$ space of Si-doped $\beta$-Ga$_2$O$_3$. We observe very weak intensity signal at the Brillouin zone (BZ) center associated to non-dispersive states likely related to Si impurities. We show that these states are quantized and that their corresponding momentum distribution curve (MDC) profiles remain restricted in $k$ space over a wide $E$ range. 

Single crystals of $\beta$-Ga$_2$O$_3$ were grown using an edge-defined film-fed growth method and electron-doped to a concentration of $\sim 1\times10^{-19}$ cm$^{-3}$ with Si atoms \cite{AidaJJAP2008}. ARPES measurements were performed with photon energy (h$\nu$) in the 20 - 200 eV range at the U1-NIM and Apple-PGM beamlines of the Synchrotron Radiation Center (Stoughton, WI) using a VG-Scienta R4000 and a VG-Scienta SES 2002 multichannel analyzers, respectively. Additional measurements were performed in Tohoku University using a VG-Scienta SES 2002 multichannel analyzer and the He-I$\alpha$ line of a He-discharging lamp (h$\nu=21.218$ eV). The samples were cleaved \emph{in situ} along the (100) plane and maintained in a vacuum better than $10^{-10}$ Torr. A previous STM study on samples from the same batch indicates that the cleaved samples have atomically flat surfaces \cite{IwayaAPL2011}. We systematically checked that the samples were not charged by varying the intensity of the incident light beam.

\begin{figure}[!t] 
\includegraphics[width=8.5cm]{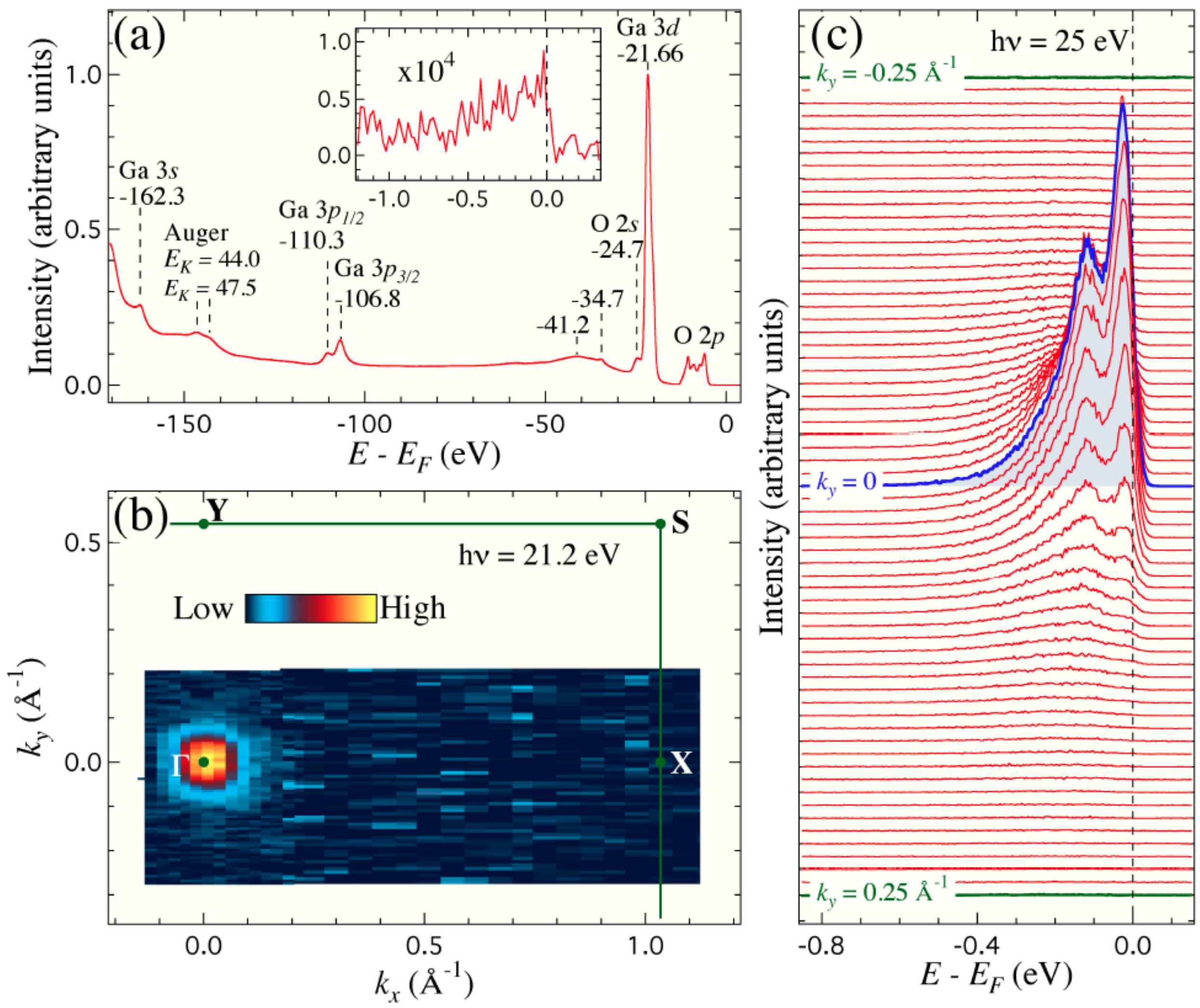} 
\caption{\label{Fig01_CoreLevels}(Color online) (a) Photoemission spectra of $\beta$-Ga$_2$O$_3$ recorded with h$\nu=195$ eV. Inset: 10$^4$ zoom near $E_F$. (b) FS intensity map. (c) EDCs recorded around $\Gamma$ with h$\nu=25$ eV.} 
\end{figure}

We show in Fig. \ref{Fig01_CoreLevels}(a) a typical core level spectrum of $\beta$-Ga$_2$O$_3$ recorded with h$\nu=195$ eV. In addition to excitations at -146.5 eV and -143 eV attributed to Auger transitions from their h$\nu$ dependence (kinetic energies $E_K$ of 44.0 and 47.5 eV, respectively), most of peaks can be simply assigned to Ga and O core levels. The spectrum is dominated by the weakly dispersive Ga 3$d$ states at a binding energy ($E_B$) of 21.66 eV. In comparison, the intensity of the valence band originating from O 2$p$ states is one order of magnitude weaker. It locates between 4.6 to 13 eV (tail to tail) below $E_F$. As expected from the 4.5-4.9 eV semiconducting gap reported for this material \cite{Tippins1965,BlasseJPCS1970,MatsumotoJJAP1974,UedaAPL97}, the spectral intensity is almost 0 up to $E_F$. Nevertheless, the near-$E_F$ zoom displayed in inset indicates a finite density of states, with a spectral intensity 10$^3$ times smaller than that of the valence band at this particular h$\nu$ value. 

This small density of states at $E_F$ was not detected in previous ARPES studies focusing on the valence band and the semiconducting gap \cite{LovejoyAPL2009,JanowitzNJP13}. To characterize it further, we performed careful ARPES measurements near $E_F$. Band structure calculations \cite{HY_HePRB2006} predict a minimum at the BZ center ($\Gamma$) for the conduction band of $\beta$-Ga$_2$O$_3$. Accordingly, the FS displayed in Fig. \ref{Fig01_CoreLevels}(b) shows only a very small spot of intensity at $\Gamma$. However, the energy distribution curves (EDCs) corresponding to an ARPES cut crossing the $\Gamma$ point given in Fig. \ref{Fig01_CoreLevels}(c) do not show evidence for dispersion within our experimental accuracy. The spectrum recorded exactly at $\Gamma$ shows a 2-peak feature that will be described below. The spectral intensity fades rapidly away from $\Gamma$ and is already virtually zero at $k_y=\pm 0.25$  \AA$^{-1}$, where no Fermi edge is detected.  

\begin{figure}[!t] 
\includegraphics[width=8.5cm]{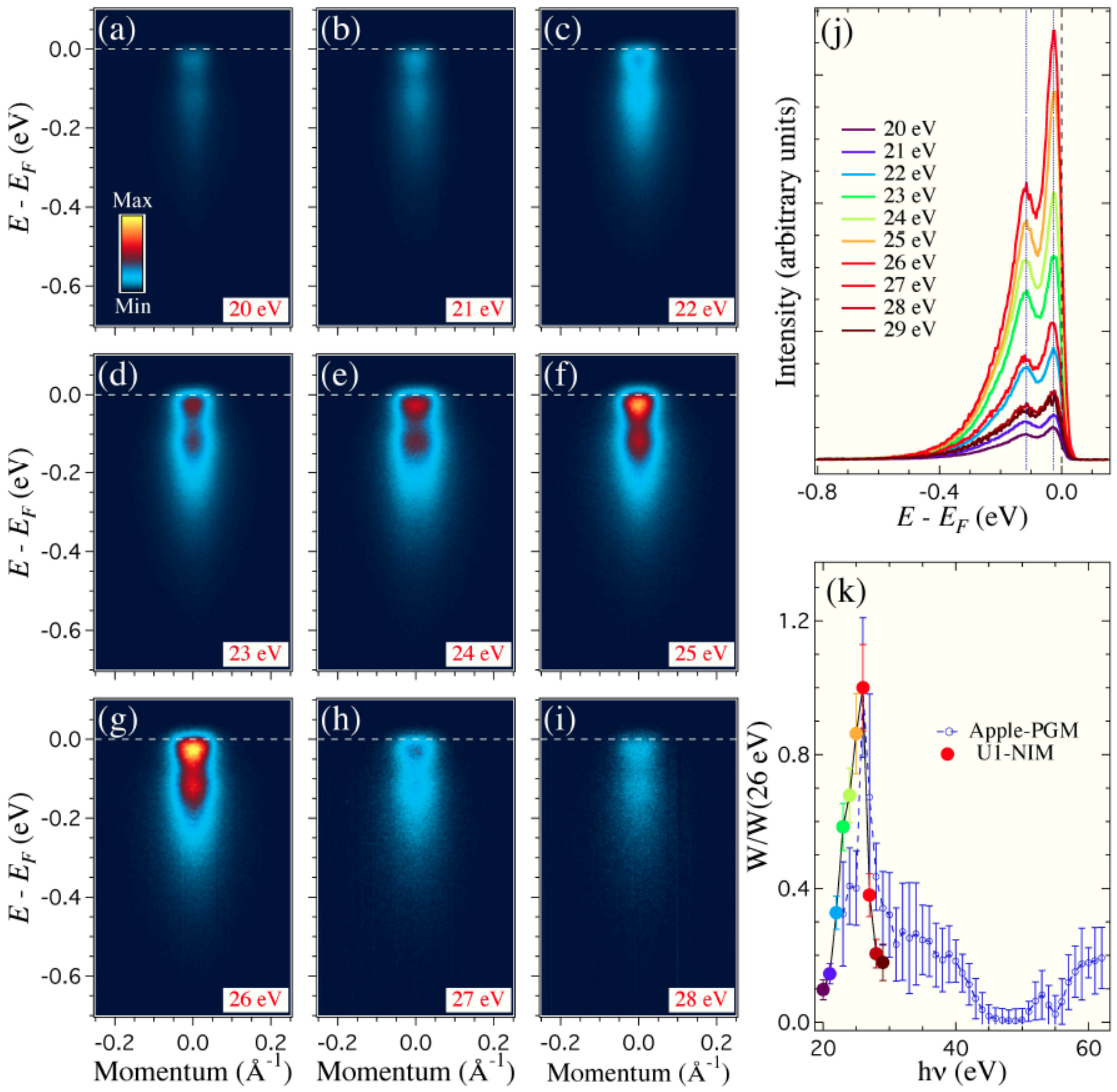} 
\caption{\label{Fig02_hv}(Color online) (a)-(i) h$\nu$ dependence of an ARPES cut crossing the $\Gamma$ point. (j) h$\nu$ dependence of the EDC at the $\Gamma$ point recorded at the U1-NIM beamline of the Synchrotron Radiation Center. The blue dot lines are guides for the eye. (k) h$\nu$ dependence of the spectral weight $W$ integrated between -0.6 to 0.1 eV and normalized at h$\nu=26$ eV. The color code for the data recorded at the U1-NIM beamline (full symbols) is the same as in panel (j).} 
\end{figure}

The relationship between $k_z$ and the h$\nu$ allows us to estimate the electronic dispersion along this direction \cite{DamascelliPScrypta2004}. Such procedure has been performed previously for the valence band of $\beta$-Ga$_2$O$_3$ \cite{LovejoyAPL2009}. To investigate the $k_z$ variations of the near-$E_F$ states, we performed h$\nu$ dependent measurements over a wide $E$ range. We show in Figs. \ref{Fig02_hv}(a)-(i) ARPES intensity cuts recorded with h$\nu$ between 20 eV and 28 eV. With h$\nu$ increasing from 20 eV, the spectral intensity increases and reaches a maximum at 26 eV, before it starts to decrease. Although the intensity of the ARPES spectra depends on h$\nu$, the spectral shape seems not to vary significantly. In particular, the 2-peak structure is well observed at each h$\nu$ value. This assertion is confirmed by the EDCs at the $\Gamma$ point, which are shown in Fig. \ref{Fig02_hv}(j). Despite strong intensity variation, the lineshape of the EDCs and the position of the two peaks are quite constant. This indicates that the states do not disperse along $k_z$, in contrast to band structure calculations suggesting that the bottom of the conduction band at $\Gamma$ is highly dispersive along $k_z$ \cite{HY_HePRB2006}. As we show in Fig. \ref{Fig02_hv}(k), we find a resonance at 26 eV that we can assigned univocally to neither Ga nor Si electronic states.

\begin{figure}[!t] 
\includegraphics[width=8.5cm]{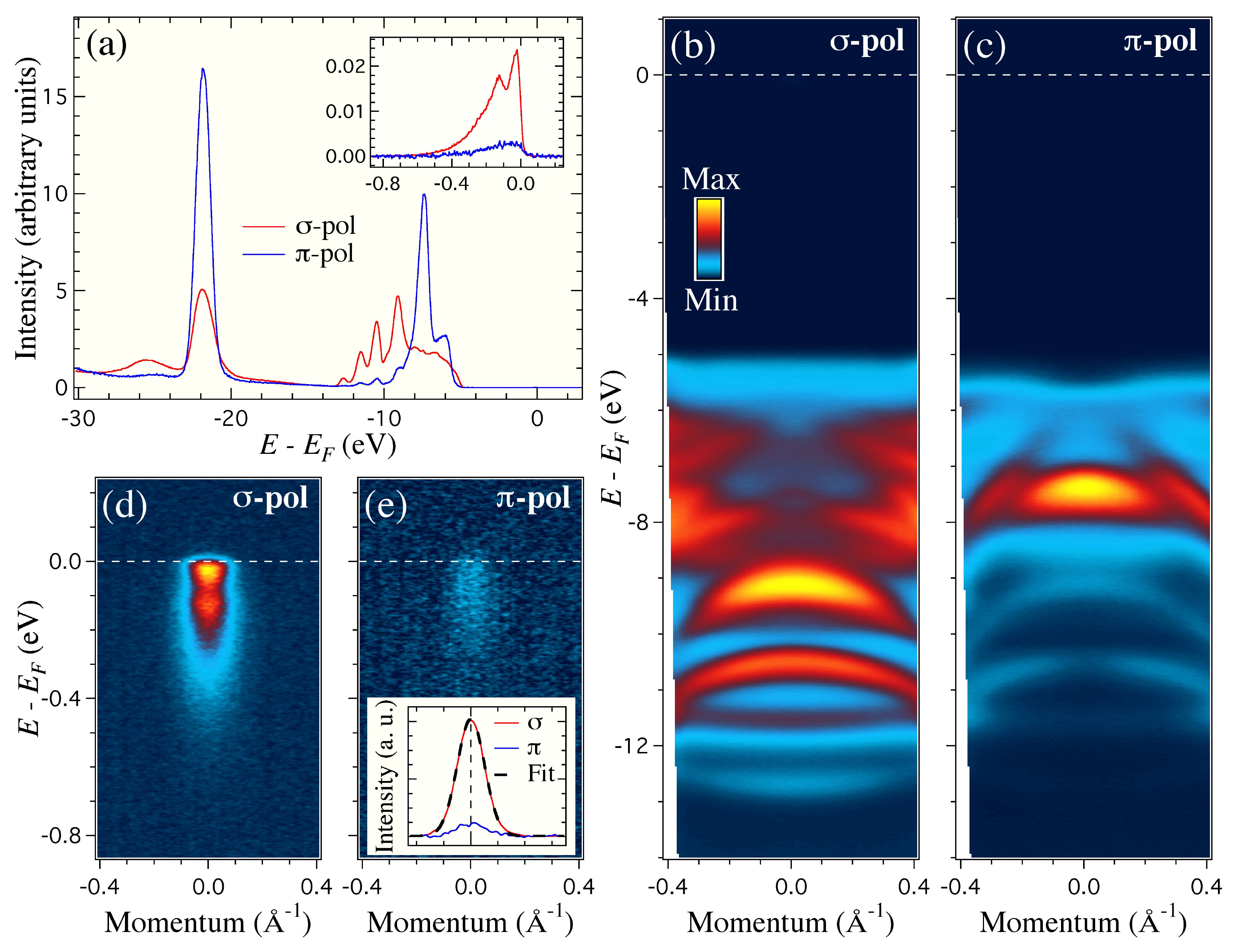}
\caption{\label{Fig03_pol}(Color online) (a) Comparison of EDCs at the $\Gamma$ point recorded using h$\nu=62$ eV with $\pi$ and $\sigma$ polarizations. An Al filter was used to suppress the strong second order light background. Inset: near-$E_F$ zoom. (b) ARPES intensity plot near $\Gamma$ recorded with 62 eV $\sigma$-polarized photons. (c) Same as (b) but using $\pi$-polarized photons. (d) Near-$E_F$ zoom of the ARPES intensity plot in (b). (e) Near-$E_F$ zoom of the ARPES intensity plot in (c). Inset: MDCs corresponding to data in (d) and (e), integrated from $E_B=170$ meV up to $E_F$. The MDC recorded with $\sigma$-polarized light is perfectly fit with a gaussian distribution.} 
\end{figure}

Additional information on the near-$E_F$ states can be obtained by varying the polarization of the incident photons. In Fig. \ref{Fig03_pol}, we compare spectra at the $\Gamma$ point obtained with $\pi$ and $\sigma$ polarized photons. Unlike the $\pi$ polarization, the $\sigma$ polarization is not pure in our experiment and the corresponding vector potential has a finite component perpendicular to the surface. Indeed, this component is mainly responsible for the differences in the EDCs displayed in Fig. \ref{Fig03_pol}(a) since they are recorded at normal emission and thus they should be identical for pure $\pi$ and $\sigma$ configurations. As confirmed with the ARPES intensity plots shown in Figs. \ref{Fig03_pol}(b) and \ref{Fig03_pol}(c), polarization strongly modulates the spectral intensity. This is also true for the near-$E_F$ states, as indicated by the near-$E_F$ ARPES intensity plots given in Figs. \ref{Fig03_pol}(d) and \ref{Fig03_pol}(e) for $\sigma$ and $\pi$ polarizations, respectively. The strong lost of spectral intensity in the $\pi$ configuration as compared to the $\sigma$ configuration, even for the normal emission EDCs shown in the inset of Fig. \ref{Fig03_pol}(a), strongly suggests that the states near-$E_F$ are associated with $z$-oriented orbitals. Due to the absence of $k$ dispersion, the states observed are most likely related with the 3$p_z$ orbitals of the Si donors. However, our experiments do not allow us to completely rule out alternative scenarios such as a surface state related to the conduction band. In this latter case though, one would expect clear in-plane dispersion, which differs from our observation.

\begin{figure}[!t] 
\includegraphics[width=8.5cm]{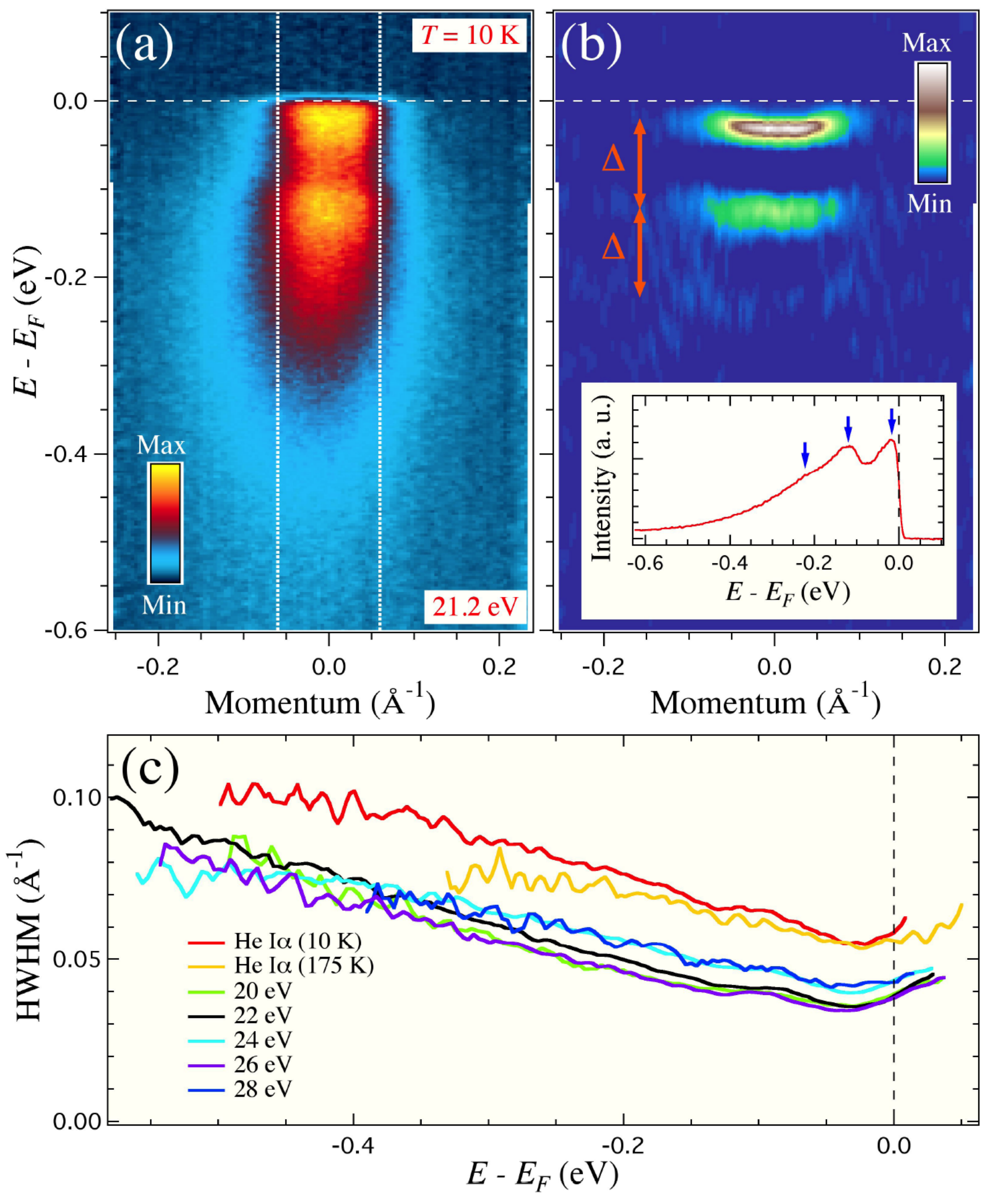} 
\caption{\label{Fig04_mode}(Color online) (a) ARPES intensity plot recorded at 10 K with the He I$\alpha$ line. (b) Corresponding curvature intensity plot \cite{Zhang_RSI2011}. Inset: EDC integrated between the vertical white dotted lines in (a). Blue arrows indicate the position of the electronic states. (c) HWHM of the MDCs as a function of $E_B$ for different h$\nu$ values. The results are from the same measurement, except for data obtained with the He discharged lamp (21.218 eV). The synchrotron data have been obtained at 50 K while the He lamp data were recorded at 10 K and 175 K.} 
\end{figure}

Our previous assignment is somewhat unconventional since an object localized in the $r$ space, such as a Si dopant atom, should lead to electronic states completely delocalized in $k$ space, which contrasts with our observation of ARPES intensity uniquely in the vicinity of the BZ center. To solve this apparent dilemma, we proceed in two steps. We first turn our attention to the lineshape associated with the near-$E_F$ states. For this purpose, we performed measurements at lower $T$ (10 K) and took advantage of the higher $E$ resolution provided by a He-discharge lamp. The resulting ARPES intensity plot is displayed in Fig. \ref{Fig04_mode}(a). The two peaks observed in the synchrotron data are still observed at $18\pm 3$ meV and $120\pm 9$ meV below $E_F$. In addition, the corresponding curvature intensity plot \cite{Zhang_RSI2011} shown in Fig. \ref{Fig04_mode}(b) reveals a third feature at $220\pm 20$ meV that is confirmed by a broad shoulder in the integrated EDC displayed in the inset of the same panel. Although it cannot be resolved within the current experiment, the tail of the EDC extends to relatively high $E_B$, suggesting the possible presence of additional peaks. As far as our measurements allow us to resolve them, the three peaks observed are evenly spaced by $\Delta\simeq 100$ meV. Noteworthy, we found experimentally that $\Delta$ varies up to 10\% from one sample to another, suggesting a possible relationship with Si doping or with the precise distribution of the Si impurities.  

What can cause such a quantization of the electronic states? One may naively propose that a bosonic mode of energy $\Delta$ interacts with the electronic structure. Indeed, Raman studies of $\beta$-Ga$_2$O$_3$ report several phonon modes for energies slightly smaller than 100 meV \cite{GaoAPL81, RaoAPL98, Y_ZhaoJRS2008}. However, such interaction usually leads to an anomaly or ``kink" in the electronic band dispersion observed by ARPES due to the opening of an energy gap. For example, this situation is often found in the superconducting state of copper-based \cite{NormanPRL79,Valla_Science1999} and iron-based \cite{RichardPRL2009} high-temperature superconductors. A more likely scenario is the presence of a quantum well (QW) \cite{MuellerPRB41}. Actually, a QW has already been proposed to explain the optical absorption spectra of minority acceptor states in the vicinity of the valence band \cite{BinetAPL77}. An important corollary to the presence of a QW associated with the electronic states of the Si impurities is that these electronic states are confined only within a typical distance $\Delta r$. The Heisenberg uncertainty principle thus implies that the same states are confined in $k$ space as well within a momentum $\Delta k$ of the order of $1/\Delta r$. 

To investigate this scenario, we analyze the MDC profile of the near-$E_F$ states. In the inset of Fig. \ref{Fig03_pol}(c), we compare the MDCs at $E_F$ recorded on the same sample using $\sigma$ and $\pi$ polarizations. Except for an intensity difference, both profiles are similar and can be fit perfectly with a gaussian function. Noteworthy, such behavior is similar to first principle calculations on oxygen-vacancy-disordered ZnO predicting $k$ space semi-localization for the O vacancy impurity sates \cite{HerngPRL105}, characterized by a loss of spectral coherence away from the zone center. Interestingly, gaussian functions can be used in $\beta$-Ga$_2$O$_3$ to fit the MDCs down to at least 500 meV below $E_F$, which is way below the 3 impurity levels identified by ARPES but still inside the tail of the near-$E_F$ feature. In Fig. \ref{Fig04_mode}(c), we show the $E$ dependence of the half-width at half-maximum (HWHM) corresponding to the fits obtained for various h$\nu$ values. Except for the low $E$ part where small anomalies are observed, the HWHM increases linearly with $E_B$ but remains finite, and no obvious influence from the probing h$\nu$ is noticed. Near $E_F$, the HWHM is about $0.040\pm 0.005$ \xspace\AA$^{-1}$ for the synchrotron data presented in Fig. \ref{Fig03_pol}(c). As with the QW spacing, the HWHM varies slightly from sample to sample. For example, the high-$E$ resolution data shown in Fig. \ref{Fig03_pol}(a) are associated with a low-$T$ HWHM of $0.055\pm 0.005$ \xspace\AA$^{-1}$, value that is quite robust against $T$. 

By rounding up the HWHM to 0.05 \xspace\AA$^{-1}$ and associating this value to $\Delta k$, we find that the corresponding spatial extension of the impurity wave function is about 20 \xspace\AA. This value is quite consistent with a previous STM study on samples from the same batch \cite{IwayaAPL2011}. Indeed, the ``size" of the impurities can be characterized by a decay length in the spectral intensity at a bias voltage corresponding to the donor states. Values ranging from 6 to 24 \xspace\AA\xspace were found for Si impurities in different locations \cite{IwayaAPL2011}. Our ARPES results confirm that these states carry some $k$ information, more precisely on the $k$ distribution of the spectral coherence. As a consequence, they can promote electrical conduction for Si concentration sufficiently high to allow the wave functions from different impurities to percolate. 

We caution that our interpretation of the multiple $E$ levels in terms of a QW is not unique. Indeed, the conduction band is expected to locate in the same $E$ range and could potentially explain some of the spectral intensity. Yet, the absence of obvious $k$ dispersion remains puzzling and further theoretical and experimental studies are necessary to solve this issue. Nevertheless, our observation of a characteristic length for the $k$-space confinement of the spectral intensity is well supported and is a step towards a better understanding of the electronic properties of $\beta$-Ga$_2$O$_3$.

\vspace{1cm}
\section*{Acknowledgements}

We are grateful to J.-P. Hu, W. Ku and X. Dai for useful discussions. This work is supported by the Chinese Academy of Sciences (grant No. 2010Y1JB6), the Ministry of Science and Technology of China (grants No. 2010CB923000, No. 2011CBA0010), the Nature Science Foundation of China (grants No. 10974175, No. 11004232, and No. 11050110422), the Japan Society for the Promotion of Sciences, and MEXT of Japan. This work is based in part upon research conducted at the Synchrotron Radiation Center which is primarily funded by the University of Wisconsin-Madison with supplemental support from facility Users and the University of Wisconsin-Milwaukee.


\begin{thebibliography}{20}
\expandafter\ifx\csname natexlab\endcsname\relax\def\natexlab#1{#1}\fi
\expandafter\ifx\csname bibnamefont\endcsname\relax
  \def\bibnamefont#1{#1}\fi
\expandafter\ifx\csname bibfnamefont\endcsname\relax
  \def\bibfnamefont#1{#1}\fi
\expandafter\ifx\csname citenamefont\endcsname\relax
  \def\citenamefont#1{#1}\fi
\expandafter\ifx\csname url\endcsname\relax
  \def\url#1{\texttt{#1}}\fi
\expandafter\ifx\csname urlprefix\endcsname\relax\def\urlprefix{URL }\fi
\providecommand{\bibinfo}[2]{#2}
\providecommand{\eprint}[2][]{\url{#2}}

\bibitem[{\citenamefont{{N. Ueda, H. Hosono, R. Waseda and H.
  Kawazoe}}(1997)}]{UedaAPL97}
\bibinfo{author}{\bibnamefont{{N. Ueda, H. Hosono, R. Waseda and H. Kawazoe}}},
  \bibinfo{journal}{Appl. Phys. Lett.} \textbf{\bibinfo{volume}{71}},
  \bibinfo{pages}{933} (\bibinfo{year}{1997}).

\bibitem[{\citenamefont{{K. Iwaya, R. Shimizu, H. Aida, T. Hashizume and T.
  Hitosugi}}(2011)}]{IwayaAPL2011}
\bibinfo{author}{\bibnamefont{{K. Iwaya, R. Shimizu, H. Aida, T. Hashizume and
  T. Hitosugi}}}, \bibinfo{journal}{Appl. Phys. Lett.}
  \textbf{\bibinfo{volume}{98}}, \bibinfo{pages}{142116}
  (\bibinfo{year}{2011}).

\bibitem[{\citenamefont{{T. C. Lovejoy, E. N. Yitamben, N. Shamir, J. Morales,
  E. G. Villora, K. Shimamura, S. Zheng, F. S. Ohuchi and M. A.
  Olmstead}}(2009)}]{LovejoyAPL2009}
\bibinfo{author}{\bibnamefont{{T. C. Lovejoy, E. N. Yitamben, N. Shamir, J.
  Morales, E. G. Villora, K. Shimamura, S. Zheng, F. S. Ohuchi and M. A.
  Olmstead}}}, \bibinfo{journal}{Appl. Phys. Lett.}
  \textbf{\bibinfo{volume}{94}}, \bibinfo{pages}{081906}
  (\bibinfo{year}{2009}).

\bibitem[{\citenamefont{{C. Janowitz, V. Scherer, M. Mohamed, A. Krapf, H.
  Dwelk, R. Manzke, Z. Galazka, R. Uecker, K. Irmscher, R. Fornari, M.
  Michling, D. Schmeissser, J. R. Weber, J. B. Varley and C. G. Van de
  Walle}}(2011)}]{JanowitzNJP13}
\bibinfo{author}{\bibnamefont{{C. Janowitz, V. Scherer, M. Mohamed, A. Krapf,
  H. Dwelk, R. Manzke, Z. Galazka, R. Uecker, K. Irmscher, R. Fornari, M.
  Michling, D. Schmeissser, J. R. Weber, J. B. Varley and C. G. Van de
  Walle}}}, \bibinfo{journal}{New J. Phys.} \textbf{\bibinfo{volume}{13}},
  \bibinfo{pages}{085014} (\bibinfo{year}{2011}).

\bibitem[{\citenamefont{{H. Aida, K. Nishiguchi, H. Takeda, N. Aota, K.
  Sunakawa and Y. Yaguchi}}(2008)}]{AidaJJAP2008}
\bibinfo{author}{\bibnamefont{{H. Aida, K. Nishiguchi, H. Takeda, N. Aota, K.
  Sunakawa and Y. Yaguchi}}}, \bibinfo{journal}{Jpn. J. Appl. Phys.}
  \textbf{\bibinfo{volume}{47}}, \bibinfo{pages}{8506} (\bibinfo{year}{2008}).

\bibitem[{\citenamefont{{H. H. Tippins}}(1965)}]{Tippins1965}
\bibinfo{author}{\bibnamefont{{H. H. Tippins}}}, \bibinfo{journal}{Phys. Rev.}
  \textbf{\bibinfo{volume}{140}}, \bibinfo{pages}{A316} (\bibinfo{year}{1965}).

\bibitem[{\citenamefont{{G. Blasse and A. Baril}}(1970)}]{BlasseJPCS1970}
\bibinfo{author}{\bibnamefont{{G. Blasse and A. Baril}}}, \bibinfo{journal}{J.
  Phys. Chem. Solids} \textbf{\bibinfo{volume}{31}}, \bibinfo{pages}{707}
  (\bibinfo{year}{1970}).

\bibitem[{\citenamefont{{T. Matsumoto, M. Aoki, A. Kinoshita and T.
  Aono}}(1974)}]{MatsumotoJJAP1974}
\bibinfo{author}{\bibnamefont{{T. Matsumoto, M. Aoki, A. Kinoshita and T.
  Aono}}}, \bibinfo{journal}{Jpn. J. Appl. Phys.}
  \textbf{\bibinfo{volume}{13}}, \bibinfo{pages}{1578} (\bibinfo{year}{1974}).

\bibitem[{\citenamefont{{H. He, R. Orlando, M. A. Blanco and R.
  Pandey}}(2006)}]{HY_HePRB2006}
\bibinfo{author}{\bibnamefont{{H. He, R. Orlando, M. A. Blanco and R.
  Pandey}}}, \bibinfo{journal}{Phys. Rev. B} \textbf{\bibinfo{volume}{74}},
  \bibinfo{pages}{195123} (\bibinfo{year}{2006}).

\bibitem[{\citenamefont{{A. Damascelli}}(2004)}]{DamascelliPScrypta2004}
\bibinfo{author}{\bibnamefont{{A. Damascelli}}}, \bibinfo{journal}{Phys.
  Scrypta} \textbf{\bibinfo{volume}{T109}}, \bibinfo{pages}{61}
  (\bibinfo{year}{2004}).

\bibitem[{\citenamefont{{P. Zhang, P. Richard, T. Qian, Y.-M. Xu, X. Dai and H.
  Ding}}(2011)}]{Zhang_RSI2011}
\bibinfo{author}{\bibnamefont{{P. Zhang, P. Richard, T. Qian, Y.-M. Xu, X. Dai
  and H. Ding}}}, \bibinfo{journal}{Rev. Sci. Instrum.}
  \textbf{\bibinfo{volume}{82}}, \bibinfo{pages}{043712}
  (\bibinfo{year}{2011}).

\bibitem[{\citenamefont{{Y. H. Gao, Y. Bando, T. Sato, Y. F. Zhang and X. Q.
  Gao}}(2002)}]{GaoAPL81}
\bibinfo{author}{\bibnamefont{{Y. H. Gao, Y. Bando, T. Sato, Y. F. Zhang and X.
  Q. Gao}}}, \bibinfo{journal}{Appl. Phys. Lett.}
  \textbf{\bibinfo{volume}{81}}, \bibinfo{pages}{2267} (\bibinfo{year}{2002}).

\bibitem[{\citenamefont{{R. Rao, A. M. Rao, B. Xu, J. Dong, S. Sharma and M. K.
  Sunkara}}(2005)}]{RaoAPL98}
\bibinfo{author}{\bibnamefont{{R. Rao, A. M. Rao, B. Xu, J. Dong, S. Sharma and
  M. K. Sunkara}}}, \bibinfo{journal}{Appl. Phys. Lett.}
  \textbf{\bibinfo{volume}{98}}, \bibinfo{pages}{094312}
  (\bibinfo{year}{2005}).

\bibitem[{\citenamefont{{Y. Zhao and R. L. Frost}}(2008)}]{Y_ZhaoJRS2008}
\bibinfo{author}{\bibnamefont{{Y. Zhao and R. L. Frost}}}, \bibinfo{journal}{J.
  Raman Spectrosc.} \textbf{\bibinfo{volume}{39}}, \bibinfo{pages}{1494}
  (\bibinfo{year}{2008}).

\bibitem[{\citenamefont{{M. R. Norman, H. Ding, J. C. Campuzano, T. Takeuchi,
  M. Randeria, T. Yokoya, T. Takahashi, T. Mochiku and K.
  Kadowaki}}(1997)}]{NormanPRL79}
\bibinfo{author}{\bibnamefont{{M. R. Norman, H. Ding, J. C. Campuzano, T.
  Takeuchi, M. Randeria, T. Yokoya, T. Takahashi, T. Mochiku and K.
  Kadowaki}}}, \bibinfo{journal}{Phys. Rev. Lett.}
  \textbf{\bibinfo{volume}{79}}, \bibinfo{pages}{3506} (\bibinfo{year}{1997}).

\bibitem[{\citenamefont{{T. Valla, A. V. Fedorov, P. D. Johnson, B. O. Wells,
  S. L. Hulbert, Q. Li, G. D. Gu and N. Koshizuka}}(1999)}]{Valla_Science1999}
\bibinfo{author}{\bibnamefont{{T. Valla, A. V. Fedorov, P. D. Johnson, B. O.
  Wells, S. L. Hulbert, Q. Li, G. D. Gu and N. Koshizuka}}},
  \bibinfo{journal}{Science} \textbf{\bibinfo{volume}{285}},
  \bibinfo{pages}{2110} (\bibinfo{year}{1999}).

\bibitem[{\citenamefont{{P. Richard, T. Sato, K. Nakayama, S. Souma, T.
  Takahashi, Y.-M. Xu, G. F. Chen, J. L. Luo, N. L. Wang and H.
  Ding}}(2009)}]{RichardPRL2009}
\bibinfo{author}{\bibnamefont{{P. Richard, T. Sato, K. Nakayama, S. Souma, T.
  Takahashi, Y.-M. Xu, G. F. Chen, J. L. Luo, N. L. Wang and H. Ding}}},
  \bibinfo{journal}{Phys. Rev. Lett.} \textbf{\bibinfo{volume}{102}},
  \bibinfo{pages}{047003} (\bibinfo{year}{2009}).

\bibitem[{\citenamefont{{M. A. Mueller, T. Miller and T.-C.
  Chiang}}(1990)}]{MuellerPRB41}
\bibinfo{author}{\bibnamefont{{M. A. Mueller, T. Miller and T.-C. Chiang}}},
  \bibinfo{journal}{Phys. Rev. B} \textbf{\bibinfo{volume}{41}},
  \bibinfo{pages}{5214} (\bibinfo{year}{1990}).

\bibitem[{\citenamefont{{L. Binet and D. Gourier}}(2000)}]{BinetAPL77}
\bibinfo{author}{\bibnamefont{{L. Binet and D. Gourier}}},
  \bibinfo{journal}{Appl. Phys. Lett.} \textbf{\bibinfo{volume}{77}},
  \bibinfo{pages}{1138} (\bibinfo{year}{2000}).

\bibitem[{\citenamefont{{T. S. Herng, D.-C. Qi, T. Berlijn, J. B. Yi, K. S.
  Yang, Y. Dai, Y. P. Feng, I. Santoso, C. S\'{a}nchez-Hanke, X.Y. Gao, A. T.
  S. Wee, W. Ku, J. Ding and A. Rusydi}}(2010)}]{HerngPRL105}
\bibinfo{author}{\bibnamefont{{T. S. Herng, D.-C. Qi, T. Berlijn, J. B. Yi, K.
  S. Yang, Y. Dai, Y. P. Feng, I. Santoso, C. S\'{a}nchez-Hanke, X.Y. Gao, A.
  T. S. Wee, W. Ku, J. Ding and A. Rusydi}}}, \bibinfo{journal}{Phys. Rev.
  Lett.} \textbf{\bibinfo{volume}{105}}, \bibinfo{pages}{207201}
  (\bibinfo{year}{2010}).

\end{thebibliography}

\end{document}